\documentclass[12pt,preprint]{aastex}

\begin{document}

\title{The Detached Eclipsing Binary KV29 and the
 Age of the Open Cluster M11}

\author{Ernest A. Bavarsad\altaffilmark{1}, Eric L. Sandquist\altaffilmark{1}, Matthew D. Shetrone\altaffilmark{2}, Jerome A. Orosz\altaffilmark{1}}
\altaffiltext{1}{San Diego State University, Department of Astronomy, San Diego, CA, 92182; {\tt eabavarsad@yahoo.com}; {\tt esandquist@mail.sdsu.edu}; {\tt jorosz@mail.sdsu.edu}}
\altaffiltext{2}{University of Texas, McDonald Observatory, HC75 Box 1337-L Fort Davis, TX, 79734; {\tt shetrone@astro.as.utexas.edu}}

\begin{abstract}

We present an extensive set of photometry and radial velocities for
the detached eclipsing binary KV 29 in the intermediate-aged open
cluster M11 (NGC 6705).  Spectroscopy shows that the system is
double-lined and all available evidence (proper motion, photometry, 
and position on the CMD) indicates it is a member of the cluster.  We 
find the period of the binary to be $4.64276\pm0.00001$ days. We 
find masses $3.604^{+0.002} _{-0.011}M_{\sun}$ and
$1.837^{+0.001}_{-0.006}M_{\sun}$, and radii
$5.392^{+0.018}_{-0.035}R_{\sun}$ and $1.656^
{+0.007}_{-0.044}R_{\sun}$ for the primary and secondary stars,
respectively. Because the primary star in the binary is rapidly
evolving and is brighter than the cluster turnoff in a color-magnitude
diagram, the measurement of its radius leads to a strong constraint on
the cluster age.  We find the age of M11 to be $222^{+2}_{-3}\pm15$
Myr, where the quoted uncertainties come from statistical errors in
the calculated masses and radii, and systematic uncertainties due to
the ambiguity of the metallicity of the open cluster and variations
within the isochrone models concerning heavy elements and convective
overshooting.

\end{abstract}


\section{Introduction}

One of the most precise ways to determine the ages of stars (besides
the Sun) involves the measurement of the masses and radii of evolved
stars in detached eclipsing binaries (DEBs).  Masses are
critical inputs for stellar models that determine the evolutionary
timescales, and accurate radii can precisely identify the evolutionary
states of the stars.  Classical age measurement methods involving the
use of isochrones in the color-magnitude (CMD) plane generally have
uncertainties (often unstated) of greater than 10\% due to continuing
difficulties in the precise measurement of distance, reddening, and
chemical composition; in theoretical uncertainties in color-$T_{\rm
  eff}$ relations and stellar interior physics; and due to degeneracies in
the effects of each of these things on the isochrones.  Fortunately,
these issues can largely be avoided using DEB systems where masses and
radii can be determined using straightforward physical principles.

Not all DEBs have characteristics that allow precise determinations of
age.  The most important requirement is that at least one component
needs to be starting to rapidly evolve in size --- stars that have
changed significantly in radius from their zero-age main sequence
value break degeneracies involving uncertainties in age, distance,
reddening, model physics, and chemical composition \citep{sou04}.
Another requirement is that both stars in the DEB must show lines in
the spectrum for radial velocity measurements and the eclipses must be
strong (preferably total) so that a system inclination can be
precisely determined.  Lastly, for precise mass and age measurements
to be made, the orbital period of the system must not be so short that
interactions between the components affect the stars' characteristics. A
star evolving off the main sequence cannot have lost or gained
significant mass, as this would critically affect the derived
age. More subtly, interactions between the stars can lead to a host of
effects from nonsphericity to modified rotation (with its own
effects).

M11 is a very rich intermediate-age open cluster, which provides us
with a heavily-populated CMD and maximizes our chances of finding
useful DEB systems.  \citet{har05} and \citet{koo07} surveyed M11 for
variable stars, detecting 5 DEBs that potentially meet the criteria
for precise age determination.  In this paper, we examine the DEB KV29
($\alpha_{2000}=18^{\mathbf{h}}51^{\mathbf{m}}00\fs13,
  \delta_{2000}=-6\degr16\arcmin37\farcs4)$, identified as an A0 star
  \citep{lindblad}.  \citet{mcn77} report that the system has a
proper motion cluster membership probability of 98\%. 
  \citet{cg14} give it a 0\% membership probability based on a radial
  velocity measurement, but this can be explained by binarity. The
photometry of the system ($V = 11.92$; M11-1139 in the tabulation
  of \citealt{ste00}) places it brighter than the cluster's turnoff,
meaning that at least one of the stars is likely to be significantly
evolved. \citet{koo07} showed that the system had eclipses with
similar depths ($\Delta V \approx 0.10 $ and 0.07 mag), meaning that
the system was likely to be double lined. There were signs of modest
ellipsoidal variation of about 0.02 mag outside of eclipse, indicating
that the stars are slightly nonspherical. This is consistent with the
relatively short period of 4.64576 d they reported.

For the purposes of our age analysis below, we will need to have some
knowledge of the cluster metallicity and reddening, although we try to
use methods that minimize their importance. In
addition, it is worth keeping previous determinations of the
cluster age in mind because systematic differences between our eclipsing
binary methods and those using CMDs can identify failings in our
understanding of the physics governing these stars. \citet{sun99}
derived log$(t_{age}) = 8.4 \pm 0.1$ by comparing $UBVI$ CMDs with
theoretical isochrones, and determined an interstellar reddening $E(B-V)
= 0.428 \pm 0.027$ and a distance modulus $(m-M)_0 = 11.55 \pm 0.10$. 
For the metal content of the stars, \citet{gaw00} derived a mean [Fe/H] 
value for M11 of $0.10\pm0.03$ by analysing high-quality spectra of 10
bright K giants. As they state, this value of [Fe/H] is consistent
with the expected value derived from the cluster's position in the
disk (6.9 kpc away from the galactic center) and the trend of increasing 
metallicity with decreasing distance from the galactic center.  \citet{santos} 
found [Fe/H]$=+0.23$ from high-resolution spectroscopy of one giant,
although it should be noted
that this value involved a metallicity-dependent correction that was
extrapolated beyond the [Fe/H] values in the original calibrating
sample. (The uncorrected value was $+0.13$.)  Recently \citet{cg14} found
[Fe/H]$=+0.10\pm0.06$ from spectroscopy of 21 cluster members, and
using that composition along with $BVI$ and $ri$ photometry derived
ages between 250 and 316 Myr (depending on the model isochrones used).
\citet{bvr14} used Str\"{o}mgren photometry to find the metallicity of
M11 to be [Fe/H]$ = -0.06\pm0.05$ from 162 F-type cluster main
sequence stars (although there is a large dispersion of 0.59 dex in individual
values), and a reddening of $E(B-V)=0.45\pm0.01$ using 332 B-A3 type
stars.

\section{Observations and Data Reduction} 

\subsection{Photometry}

M11 was observed on 37 nights from June 2009 to August 2011 (with one
additional night in 2013) in $BVR_C$ filters using the 1.0m telescope
at Mount Laguna Observatory (hereafter, MLO) with the CCD camera field
covering an area approximately $13\farcm5 \times 13\farcm5$, for a
scale of about $0\farcs4$ pixel$^{-1}$.  The nights of observation are
listed in Table \ref{photobs}.  Exposure times for varied between 90
and 300 s depending on filter and atmospheric conditions.  The
exposure time was adjusted throughout the night in order to achieve a
peak of about 30,000 counts for the binary system per image.  The
seeing from our observations generally produced a FWHM of stellar
images of $1.5-2.5$ arcsec.  Image processing (overscan
subtraction, bias subtraction, and flat-field correction) was done
using standard tools in IRAF\footnote{IRAF is distributed by the
  National Optical Astronomy Observatory, which is operated by the
  Association of Universities for Research in Astronomy, Inc., under
  cooperative agreement with the National Science Foundation.}.

M11 is a heavily-populated open cluster, and crowding in the field is
significant. In order to derive precise light curves, we employed
point-spread function (PSF) photometry using DAOPHOT and ALLSTAR
\citep{daophot}.
The PSF was iteratively fit and allowed to vary across the frame
according to a quadratic polynomial. Typically around 150 relatively
isolated stars were used to determine the PSF.

After photometry was derived for each frame using ALLSTAR, we applied
ensemble methods \citep{honey,s1082} to correct the photometric
zeropoints for effects such as nightly variations in atmospheric
transparency. Because thousands of stars could typically be measured
on each image, the zeropoints are very well determined from the
stellar ensemble. As the zeropoints are refined, the median
  instrumental magnitudes for the stars become better determined as
  well. The median magnitudes and frame zeropoints were
iteratively recalculated until all values converged to better than
0.0005 mag.

\begin{figure}
\vspace{-160pt}
\hspace{-50pt}
\includegraphics[scale=0.9]{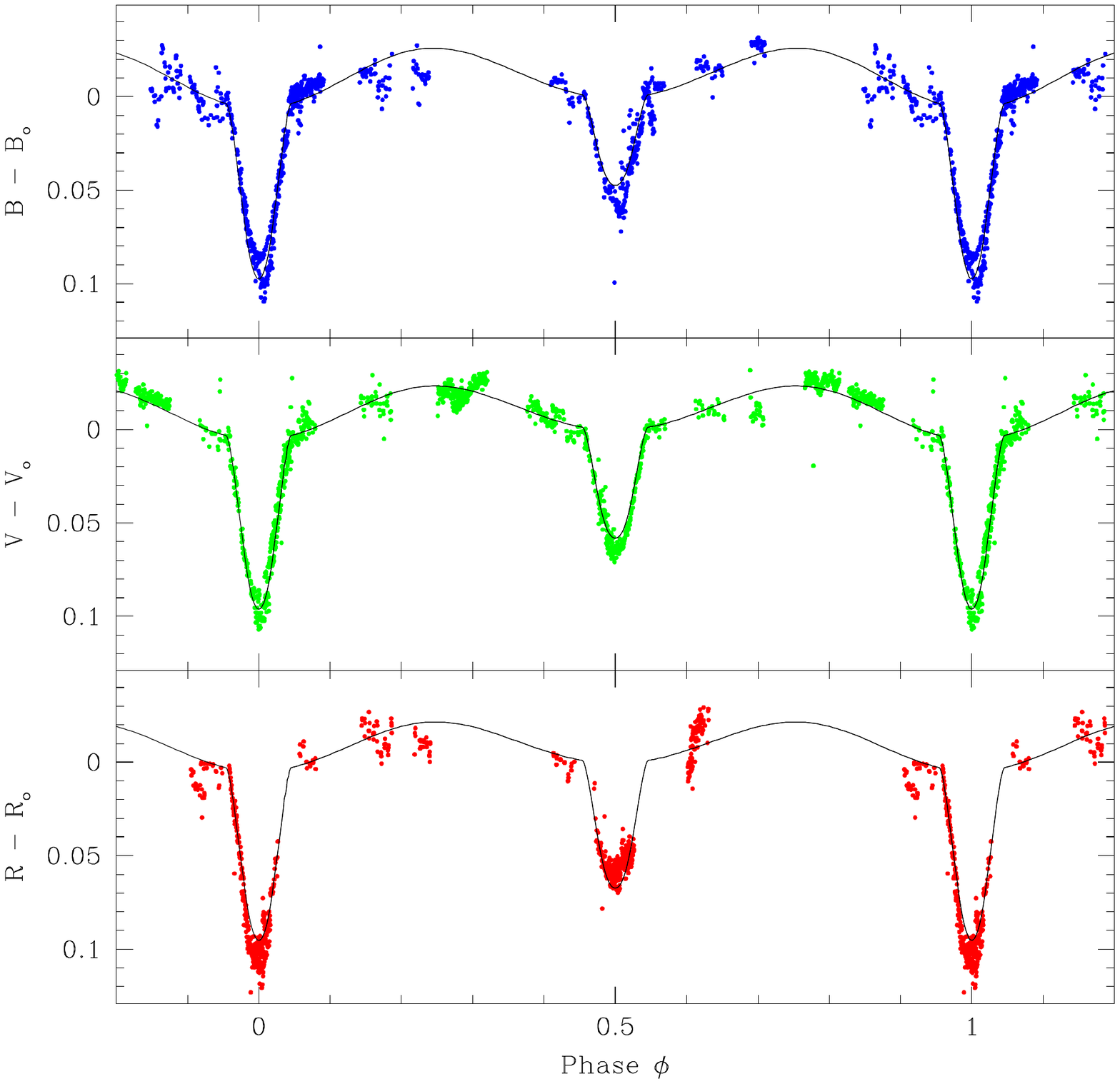}
\vspace{-140pt}
\caption{Phase-folded light curves for KV29 in $B$ ({\it top panel}),
  $V$ ({\it middle panel}), and $R_C$ ({\it bottom panel}), referenced
  to the median magnitude in each filter.  The model fit (black line) is 
  from the full data run with fixed limb darkening coefficents. \label{phot}}
\end{figure}

\begin{figure}
\vspace{-160pt}
\hspace{-50pt}
\includegraphics[scale=0.9]{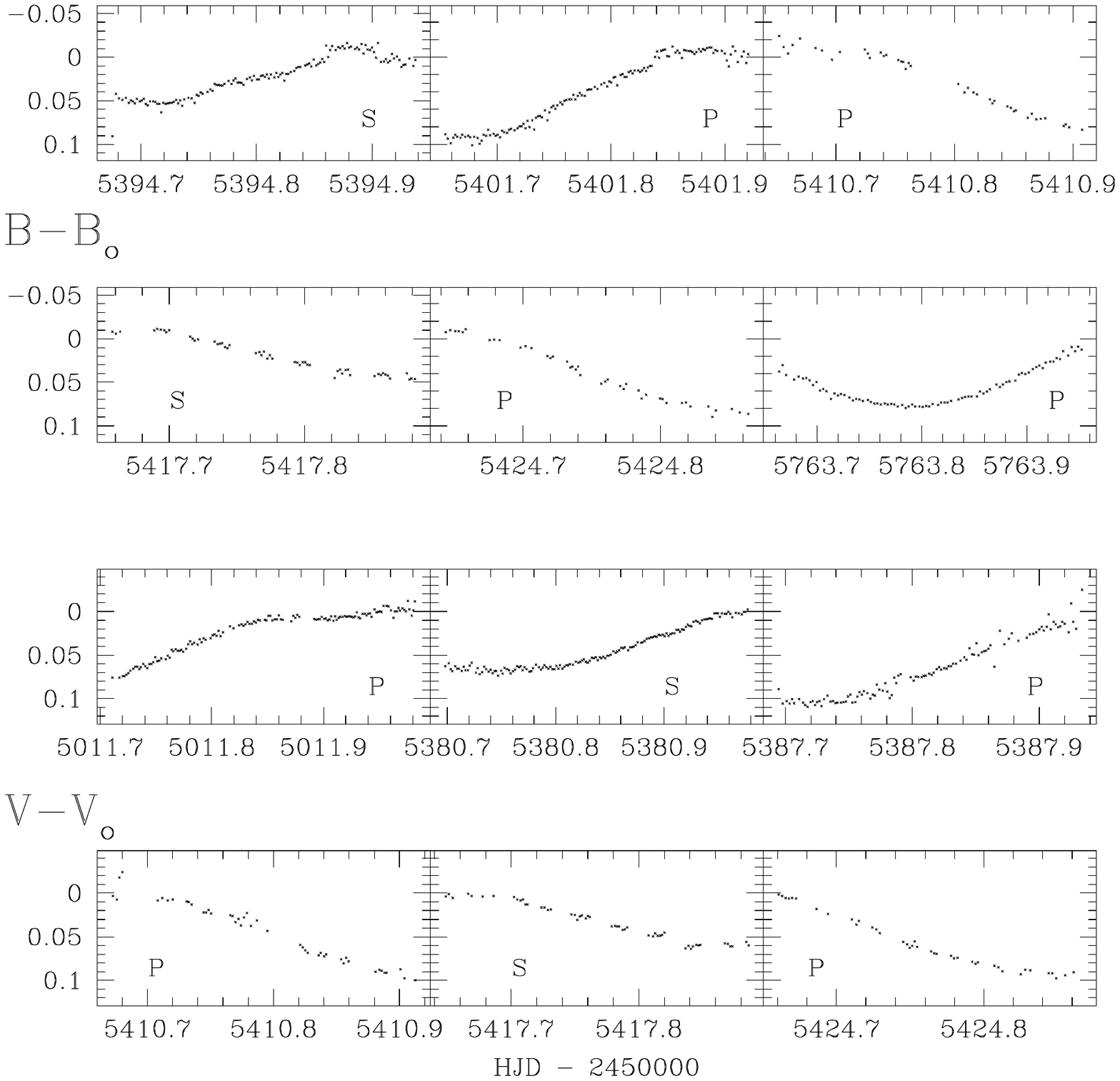}
\vspace{-140pt}
\caption{MLO photometry of all eclipse nights in $BV$ filters,
  referenced to the median magnitude.  Graphs are labeled ``P'' and
  ``S'' for primary and secondary eclipses,
  respectively. \label{panelBV}}
\end{figure}

\begin{figure}
\vspace{-160pt}
\hspace{-50pt}
\includegraphics[scale=0.9]{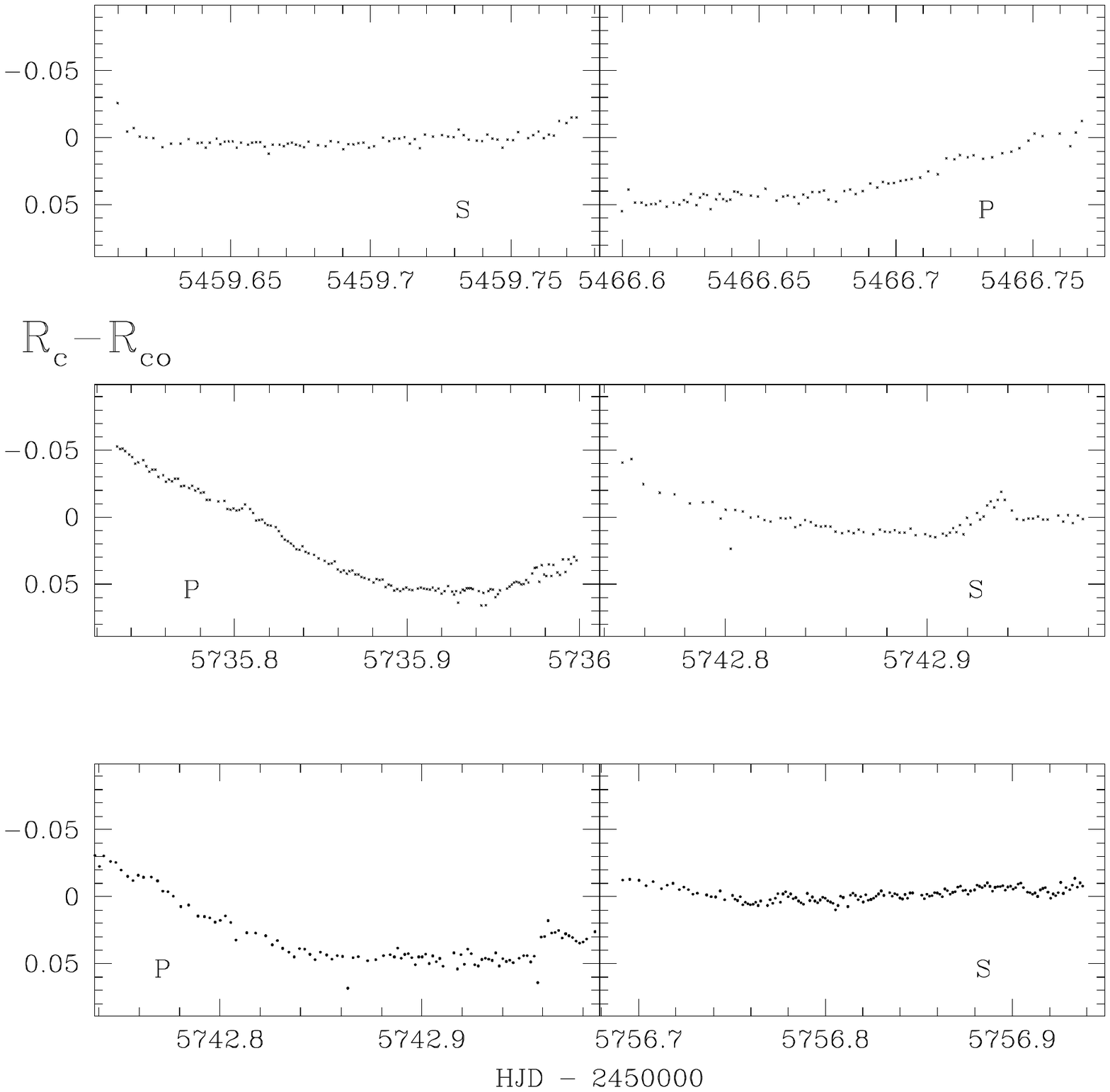}
\vspace{-140pt}
\caption{MLO photometry of all eclipse nights in $R_c$, referenced to
  the median magnitude.  Graphs are labeled ``P'' and ``S'' for
  primary and secondary eclipses.\label{panelR}}
\end{figure}

The depth of the primary eclipses are fairly consistent from filter to filter, 
but the secondary eclipse is deepest in $R_c$ ($\Delta R_c = 0.07$) as 
shown in Fig. \ref{phot}. The radial velocities
(see \S \ref{spect}) show that the cooler, less massive star is behind its
companion at the time of secondary eclipse.
Figures \ref{panelBV} and \ref{panelR} show the photometry taken on
individual nights of eclipse. Most importantly, the secondary eclipses
show evidence of being flat at minimum (see HJDs 2455380, 2455742, and
2455756), suggesting the eclipses are total. This makes it easier to
disentangle the amount of light contributed by each star.

The light curves often show variations that may be independent of
  the eclipses. While spots are not generally expected for stars with
  radiative envelopes, observations from the {\it Kepler} mission
  indicate that they do occur. \citet{balona} found evidence of
  activity among about 40\% of A-type stars. The amplitude of the
  variations was most commonly in the tens of parts per million, but
  did occasionally reach mmag levels.  Because the primary star is
  more than 3 mag brighter than the secondary star, an astrophysical
  source of variability almost certainly originated on the
  primary. However, longer term monitoring is needed to establish
  the nature of the variability.

\subsection{Spectroscopy}\label{spect}

Our spectra were obtained at the Hobby-Eberly Telescope (HET) with the
High Resolution Spectrograph (HRS; \citealt{tull98}) as part of normal
queue-scheduled observing \citep{shet07}.  The configuration of the
HRS was chosen based upon the spectral line widths and strength of the
secondary in the first spectrum taken of each object. KV29 was
  observed with a resolution $R=30000$, a $2\arcsec$ fiber feed, and
  binning to approximately 2.7 pixels per resolution element.
The configuration covers 4825 \AA\ to 6750
\AA\ with a small break at 5800 \AA\ between the red and blue CCDs,
although we used only the blue section of the spectra.  Exposure times
were between 480 and 700 s.  The data were reduced using the echelle
package within IRAF for fairly standard bias and scattered light
removal, 1D spectrum extraction, and wavelength calibration.

We obtained a total of 17 spectra between May 2010 and September
2011. In two of these spectra, the lines were heavily blended
(to the point that cross correlation peaks were fully blended) and two
others had moderate blending. However, based on the width of the
lines, the brighter star is clearly rotating more rapidly
than the secondary, meaning that the spectral features of the
secondary could be marginally identified in the spectra with the
heaviest blending.

Our radial velocity measurement procedure made use of broadening functions
\citep{bfs,bf02} and spectral disentangling \citep{gonz}.  Broadening
functions (BFs) transform a sharp-lined spectrum into a broadened
spectrum through a convolution, and they contain information on the
Doppler shifting of the spectrum as well as broadening effects like
rotation. BFs generally improve the measurement of Doppler shifts
relative to cross correlation methods in the presence of substantial
rotational broadening and overlapping spectral lines, and this is of
great importance for a short-period binary like KV 29.  We resampled
to a common logarithmic wavelength spacing at approximately the same
pixel resolution as the original spectra, and employed IDL routines
provided publicly by S. Rucinski\footnote{\tt
  http://www.astro.utoronto.ca/$\sim$rucinski/SVDcookbook.html} to
derive the broadening function via singular value decomposition.  We
used different synthetic spectrum templates with temperatures $T_{\rm
  eff} = 10000$ and 8500 K for the two stars, but both having $\log g
= 3.557$, and [Fe/H]$=+0.10$. We fit the peaks in the BFs with rotational
broadening profiles assuming 50 and 20 km s$^{-1}$,
allowing for a modest amount of variation around those values (15 and
4 km s$^{-1}$, respectively). We found that restricting the analysis
to wavelengths between 5000 and 5400 \AA\ produced the clearest
broadening functions because this range contained the strongest and
best-measured spectral lines. 

To determine precise radial velocities even for the spectra with the
most heavily blended lines, we used spectral disentangling.
Disentangling involves an iterative procedure of redetermining radial
velocities and averaged spectra for each component of the binary
repeatedly, with subtraction of the averaged spectrum of one star from
each collected spectrum during the measurement of the radial
velocities and averaged spectrum (shifted to the rest frame) of the
other star. By removing the companion star's spectrum as best we
possibly can, we minimize systematic effects from line blending on the
measured radial velocities. The technique works best when there are a
sufficient number of spectra taken at different orbital phases with
good velocity separation. In our case, we used the 9 spectra with the
largest separation in the determination of the average spectra, and
the iterations converged rapidly.

\begin{figure}
\vspace{-160pt}
\hspace{-50pt}
\includegraphics[scale=0.9]{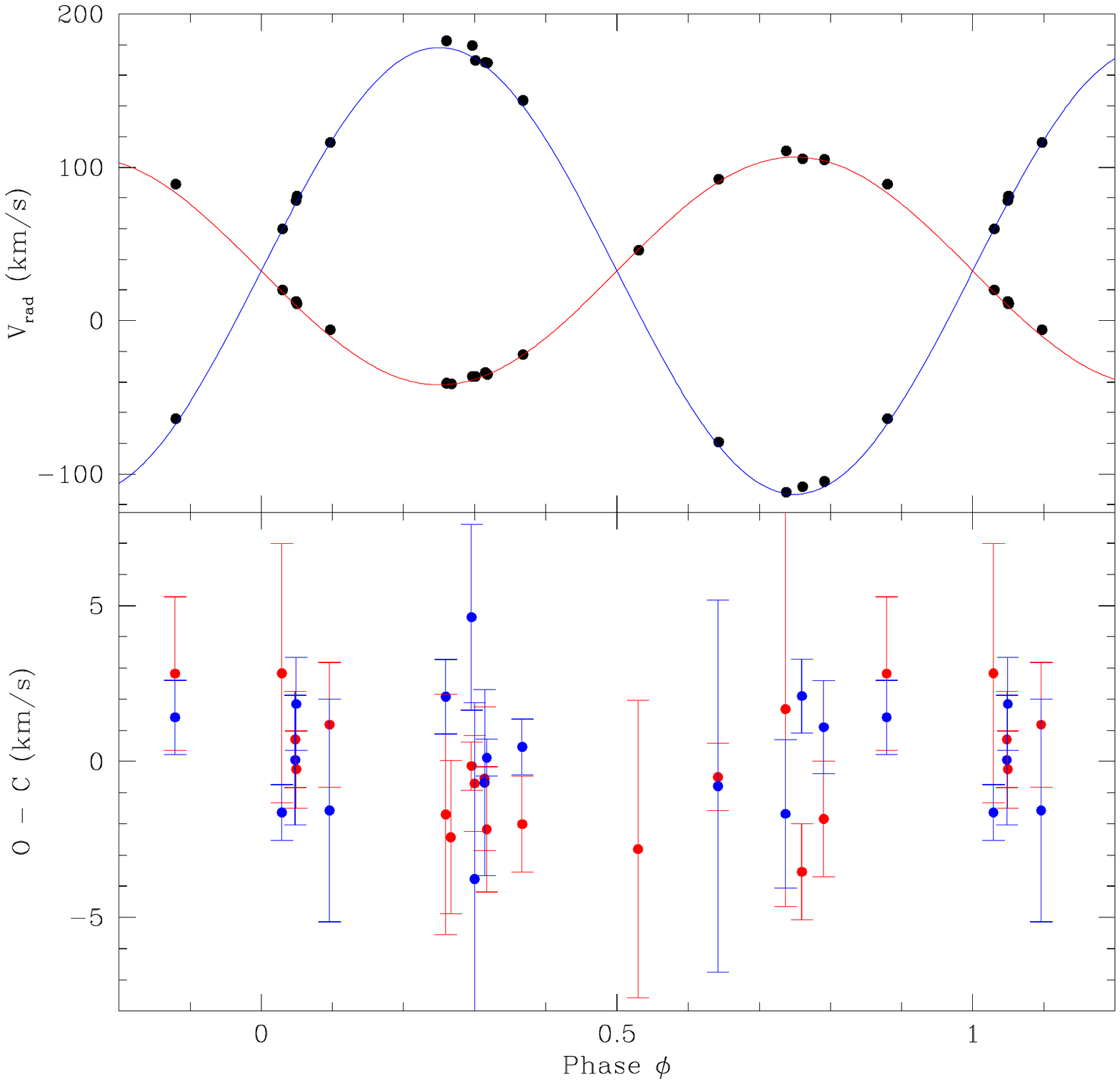}
\vspace{-140pt}
\caption{Radial velocities for KV29 phased to the binary orbit. The colored 
lines are the best fitting models.  The bottom plot is an O-C 
graph with error bars scaled to give a reduced $\chi^2 \approx 1$.\label{OC}} 
\end{figure}

We corrected the measured values for heliocentric velocity and
spectrograph zero-point offsets. Radial velocity standards were
observed on most nights that the binary was, and the difference
between the measurement and literature value was used as a
zero-point correction for the eclipsing binary measurements. If a
standard was not observed on the same night, the correction was
determined by averaging measurements from nights nearest in time. From
experience, the radial velocity corrections vary only slightly (by a few
hundred m s$^{-1}$) from night to night, but do vary significantly
with instrument configuration and season. The final radial velocity
values are given in Table \ref{specobs} and plotted (along with an
$O-C$ diagram) in Fig. \ref{OC}.

By comparing synthetic spectra with the disentangled spectra, we
derived constraints on the rotational velocities and temperatures of
the two stars. We find temperatures of approximately 10000 and 7500 K,
although we will hone our temperature estimates in the next section
using photometry. In any case, the spectral types appear to be A0IV
and A6V. (Before it reached the turnoff and evolved significantly, the primary
probably had a spectral type of B9V.) We find rotation speeds of 46
and 18 km s$^{-1}$ with uncertainties of approximately 4 and 2 km
s$^{-1}$, respectively. Based on these speeds and later determination
of the stellar radii, both stars appear to be rotating approximately
in synchronism with the orbit. This is somewhat surprising given that
the primary star probably expanded by about a factor of two in the
last 40 Myr as a result of its evolution. The rotation of the primary
may be lagging behind synchronism slightly --- further analysis of
this system could place interesting constraints on tidal effects on
the rotation.

\section{Binary Star Modeling}

In order to model the photometric and spectroscopic data of KV29, we
used the ELC code \citep{oro00}, which employs a Markov chain Monte
Carlo \citep{teg04} algorithm to optimize the fit.
Because the spectroscopic results indicate the binary orbit has zero
eccentricity, some orbit parameters (namely the velocity
semi-amplitude of the primary star $K_p$, and the ratio of the stellar
masses $Q=M_2/M_1$) can be determined from spectroscopic velocities
independent of the light curves. We therefore fitted the radial
velocities first, searching for the minimum $\chi^2$, while also
examining models with alternate parameter sets in order to evaluate
uncertainties. The uncertainties in the binary model parameters were
derived from the range of parameter values that produce a total
$\chi^2$ within 1 of the minimum value, which approximates a $1\sigma$
uncertainty \citep{avi76}.  Because the velocity measurement
uncertainties are used in the calculation of $\chi^2$, it is important
that these uncertainties be as realistic as possible. Usually
measurement uncertainties calculated during the data reduction process
are underestimated, which would inflate the total $\chi^2$ value and
lead to an underestimation of the uncertainties in the binary star
model parameters. With awareness of this, we scaled the uncertainty
estimates of the radial velocities from the spectroscopic analysis
(and later, the photometric measurements in each filter) upward to
return a reduced $\chi^2$ value of 1 for each measured quantity in
order to be consistent with the observed scatter around the best fit
model.  The results for $K_p$ and $Q$ are tabulated in Table
\ref{constraints}.

\begin{figure}
\vspace{-160pt}
\hspace{-50pt}
\includegraphics[scale=0.9]{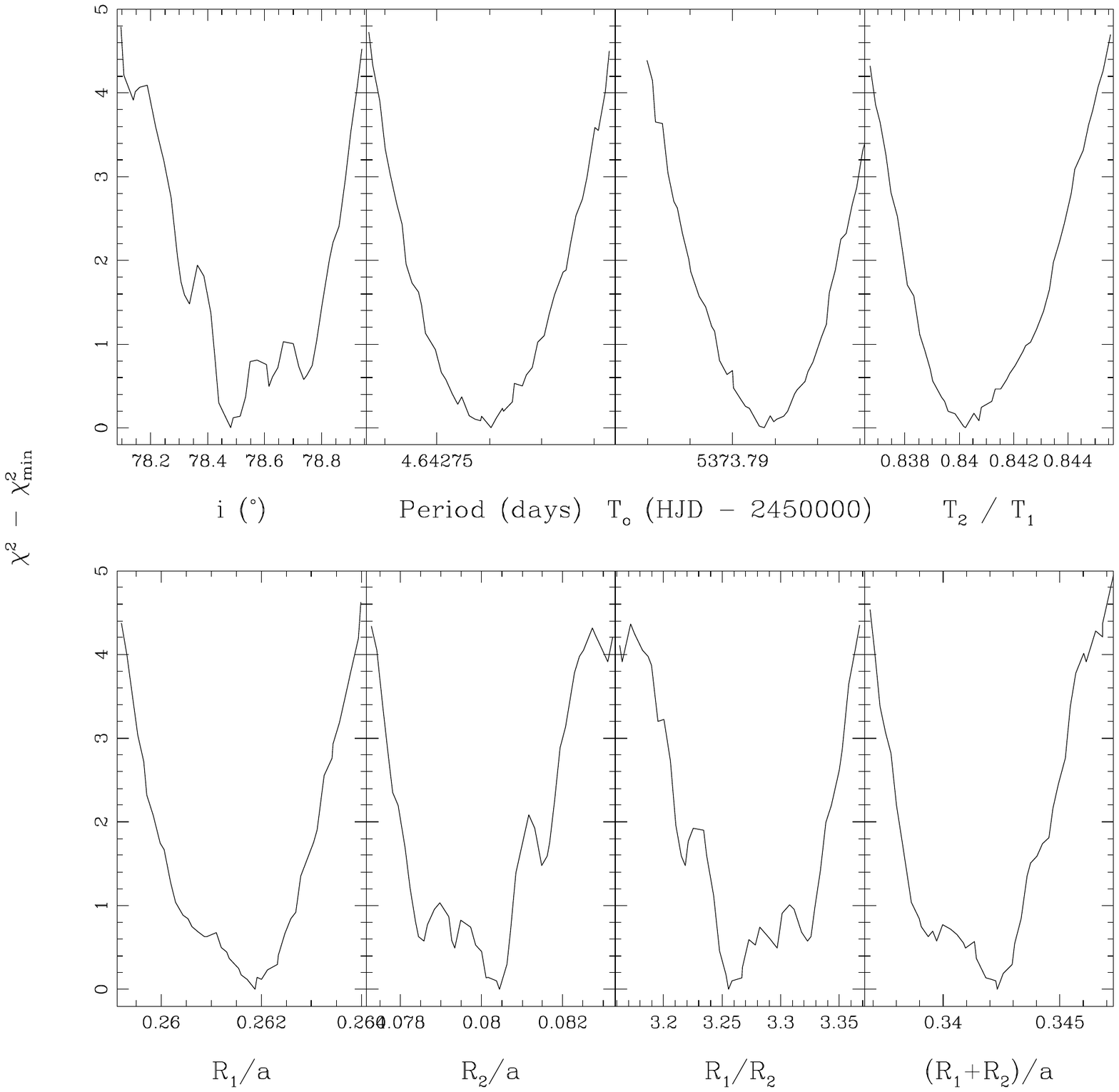}
\vspace{-140pt}
\caption{$\chi^2 - \chi^2_{min}$ plot for each of the six parameters 
in the constant limb darkening run.\label{chi2}}
\end{figure}

\begin{figure}
\vspace{-160pt}
\hspace{-50pt}
\includegraphics[scale=0.95]{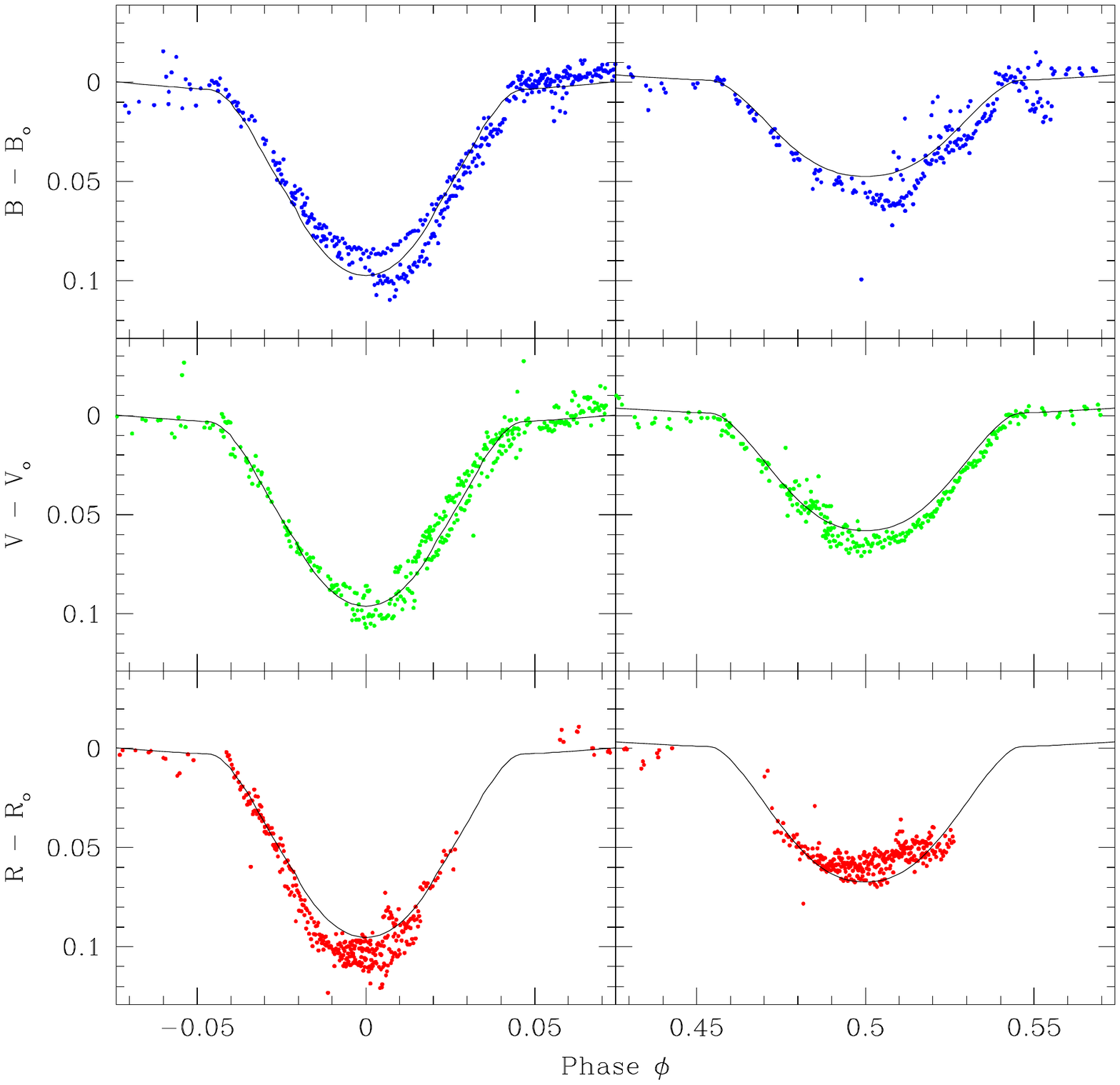}
\vspace{-140pt}
\caption{Zoom on the eclipses in $BVR_c$.  The solid lines are the best 
fitting model derived from the constant limb darkening run.  The median 
magnitude for each color has been subtracted off.\label{eclipse}}
\end{figure}

We then proceeded to light curve fitting. Because the degree of limb
darkening has a significant effect on measured radii for systems like
KV29 having shallow eclipses, we probed the potential systematic errors
by fitting with different limb darkening descriptions.  We first used a
quadratic limb darkening law with coefficients taken from \citet{cb11}
and using PHOENIX model atmospheres \citep{hau97} to normalize the
brightness from the centers of the visible star disks.
The parameters we fit for were orbital period $P$, time of conjunction
(primary eclipse) $t_o$, inclination $i$, ratio of the stellar radii
to average orbital separation $R_1/a$ and $R_2/a$, and temperature
ratio $T_2/T_1$. $K_p$, $Q$, and $T_1$ were held as external
constraints.  These observational constraints impose a $\chi^2$
penalty on models if the parameter values deviate from values
determined earlier.  The $\chi^2 - \chi^2_{min}$ plot for the 6
parameters $P$, $T_o$, $i$, $R_1/a$, $R_2/a$, and $T_2/T_1$ in this
run (constant limb darkening coefficients) can be seen in
Fig. \ref{chi2}.  A model fit of the eclipses can be seen in
Fig. \ref{eclipse}.

We also conducted fits while allowing the limb darkening coefficients
to vary, following the algorithm of \citet{kip13}.  Due to the scatter
in the observed light curves, we examined the effects of using
weighted average photometry for phase bins. (See \citealt{sm88} for a
similar usage for the short-period eclipsing binary DS Andromedae.) We
used phase bins of 0.002 in and near eclipses (0.90 to 0.10 and 0.40
to 0.60) and 0.01 elsewhere.  The uncertainty of each binned point was
taken to be the error of the weighted mean.  Results from both runs
can be seen in Table \ref{trials}, and some fitted quantities can be
seen to deviate by several standard deviations ($T_{rat}$).  The
difference in the lightcurve shapes from filter to filter has made it
difficult to fit for limb darkening coefficients, but the fitted
parameters for the binned run mostly agree with those of the full data
run.  Although there are changes to the measured stellar radii
depending on the method of fitting, it should be noted that these
small differences negligibly affect our age determination in the
next section because the primary star is in a phase of rapid radius
evolution. The masses are more important for the age determination
here, and the quantities $a \sin i$, $M_1 \sin^3 i$, $M_2 \sin^3 i$
from each model agree to less than $1\sigma$.

Ellipsoidal variations are visible in the out-of-eclipse observations
plotted in Fig. \ref{phot}, suggesting that at least one of the stars
is tidally distorted, in agreement with the expections for the short
orbital period and evolved state of the primary star. The ELC
  code uses a Roche lobe geometry to model the non-spherical nature of
  the stars in the binary and its effects on the light curve. The
Roche lobe filling factor for the primary star was calculated to be
0.470, much larger than that of the secondary at 0.176. Because the
primary star is much brighter than the secondary star, almost all of
the out-of-eclipse variation in the lightcurves is caused by the
primary.

\begin{figure}
\vspace{-160pt}
\hspace{-50pt}
\includegraphics[scale=0.9]{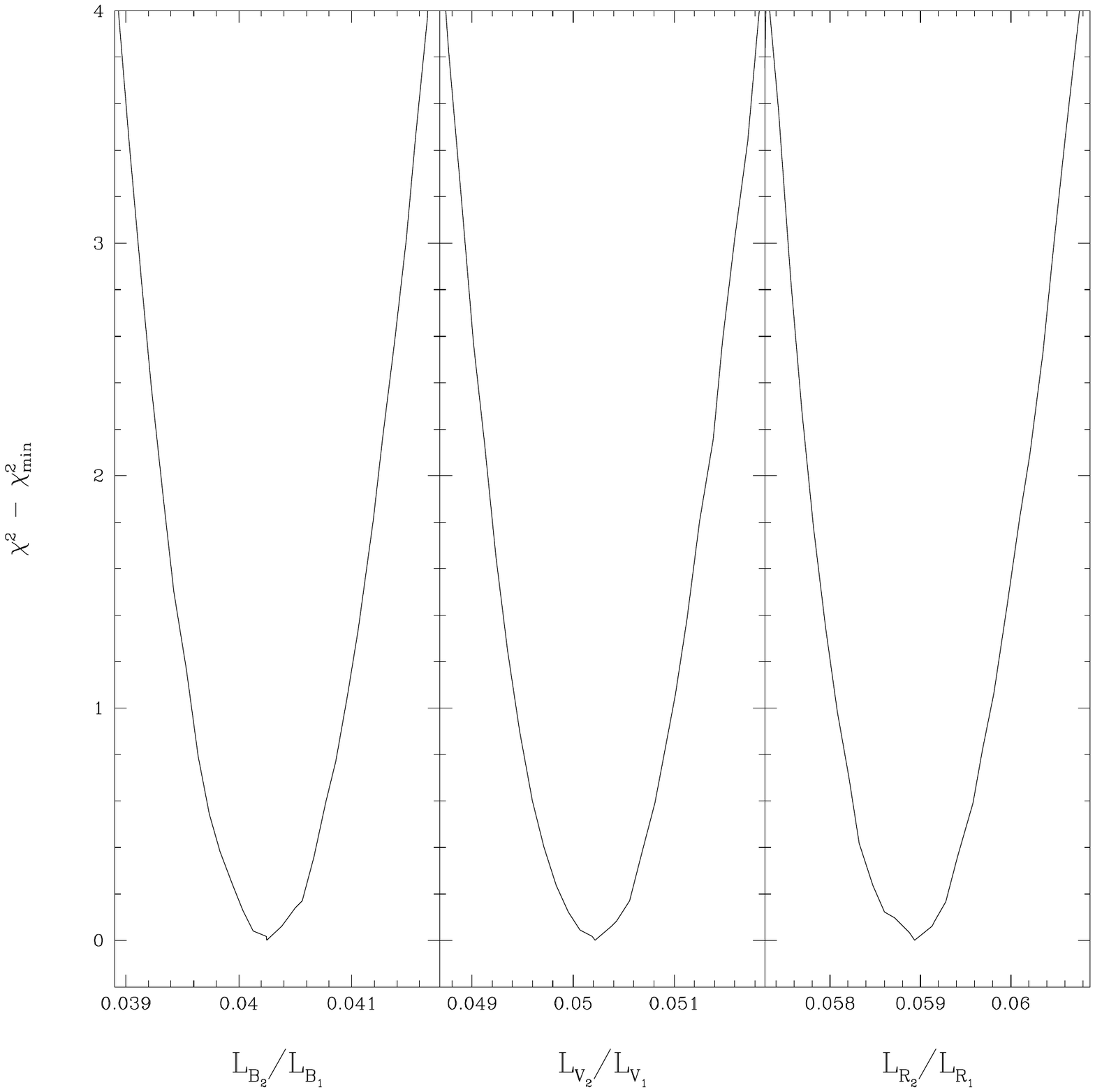}
\vspace{-140pt}
\caption{$\chi^2 - \chi^2_{min}$ plot for the luminosity ratios $(L_{2}/L_{1})$ in $BVR$. \label{colorsfig}}
\end{figure}

As mentioned before, we see that during the secondary eclipse the
primary star is totally blocking the light of its companion.  Because
the primary star alone is visible then, the secondary eclipse depths
allow us to disentangle the photometry of the two stars.  We did this
with two different methods.  First, we carefully measured the
secondary eclipse depths and used them to infer the brightness of the
primary star.  The following nights (in HJD) were used: 2455394,
2455401, 2455410, 2455417, 2455424, and 2455763 in $B$, and 2455011,
2455380, 2455387, 2455410, 2455417, and 2455424 in $V$.  We took the
average of the out-of-eclipse observations for a given night and
subtracted it from the average of the points during totality.  We
found eclipse depths $\Delta B = 0.055 \pm 0.002$ and $\Delta V =
0.066 \pm 0.002$. For stars with significant non-sphericity, the
  depths only provide upper limits to the luminosity ratio because we
  are looking down the long axis of an elongated star during the
  eclipses. We therefore derived luminosity ratios from ELC model
fits to all of the light curve data (see Fig. \ref{eclipse}). We found
the $\chi^2 - \chi^2_{min}$ for the luminosity ratios $L_2/L_1$ for
each color in the fitted limb darkening run (Fig. \ref{colorsfig}) and
used those to decompose the photometry of the stars.  The ELC models
yielded $L_{B2}/L_{B1} = 0.040\pm0.001$, $L_{V2}/L_{V1} =
0.050\pm0.001$, and $L_{R2}/L_{R1} = 0.059\pm0.001$, where the
  quoted uncertaintied are statistical. We have chosen to include a
  contribution to the uncertainty of 0.002 from systematic error
  resulting from variations in eclipse depths. The decomposed
photometry for the two stars can be seen in Table \ref{colors}.
 
Using the decomposed $B-V$ colors for both stars, we calculated
photometric $T_{\rm eff}$ values using the transformation equations of
\citet{cas10}. For that purpose, we used a cluster reddening
$E(B-V)=0.43$ \citep{sun99} and metallicity [Fe/H] $= +0.10$
\citep{gaw00}. By this method, $T_1$ was found to be $9480\pm550$ K
and $T_2 = 7810\pm480$ K.  This value for $T_2$ is roughly consistent
with what can be derived from the value of $T_1$ and the temperature
ratio derived from the fitted limb darkening run  ($T_2 \approx 8100$ K).  
Uncertainties in the effective temperatures come from uncertainty in the 
reddening and metallicity, and in the temperature ratio provided 
by ELC.  Abundance uncertainties (such as the difference between the 
\citeauthor{gaw00} spectroscopic results and the \citet{bvr14} 
determination of [Fe/H] $=-0.06$ using Str\"{o}mgren photometry) 
have negligible effects ($\sim10$ K) on this calculation.

\section{Discussion and Analysis} \label{discussion}

The radii of evolved stars can make excellent age
indicators because they can be measured to high precision in eclipsing
binary systems and their use avoids systematic errors that are commonly
present in other indicators.  However, the translation from stellar radius
to age requires the use of theoretical mass-radius isochrones.
Every isochrone set employs an assumed value for the heavy element content of
the Sun ($Z_{\sun}$) because it is not a directly measured quantity.
Because the metallicity ([Fe/H], and therefore $Z$) of other stars is
judged relative to the Sun, this has a potentially significant effect
on the age determination if the Sun's metal content is systematically
different from what is assumed.  After a thorough
re-examination of the solar abundance mix, \citet{asp09} found
$Z_{\sun} = 0.0134$, which is lower than most values assumed in
models. If the \citeauthor{asp09} value is correct, we should select
the heavy element abundance $Z$ of the models based on the revised
$Z_{\odot}$.  Using this new value of $Z_\sun$ and the \citet{gaw00}
spectroscopic [Fe/H] value, the M11 metal content should be $Z =
0.0161$. While the PARSEC models allow direct input of a particular
$Z$ value, we had to choose $Z$ values as close as possible to that of
M11 for the Victoria-Regina and BaSTI isochrone models.  The
PARSEC isochrones \citep{bre12} will be used as our primary theoretical 
isochrone set, with the Victoria-Regina \citep{vdb06} and BaSTI \citep{pie04} 
models as comparisons.

\begin{figure}
\vspace{-100pt}
\hspace{-80pt}
\includegraphics[scale=0.98]{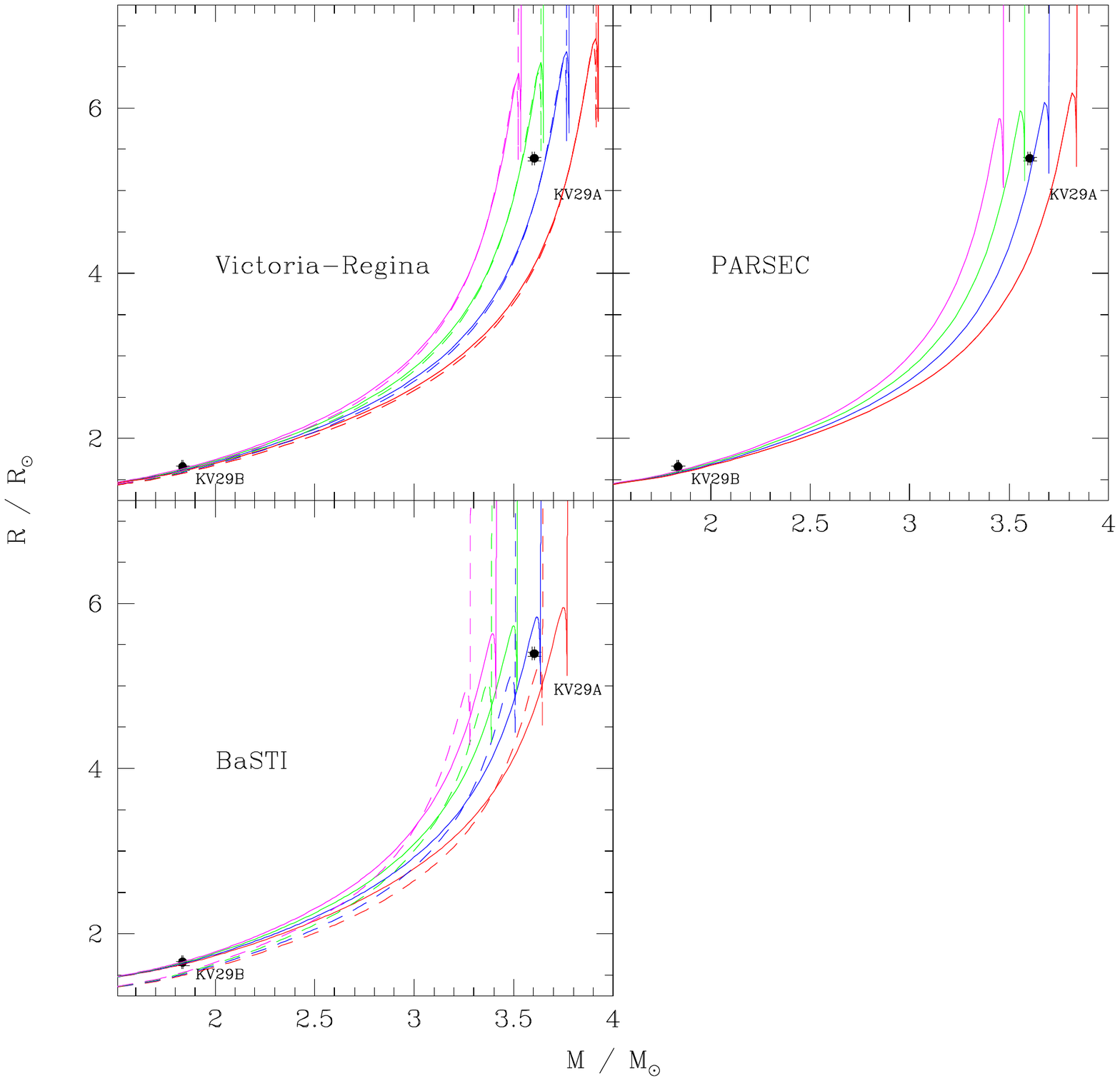}
\vspace{-150pt}
\caption{Mass-radius plots for the components of KV29.  Models have
  ages 200 (red), 220 (blue), 240 (green), and 260 Myr (magenta).  The
  PARSEC isochrones employ $Z=0.0161$.  The Victoria-Regina isochrones
  use $Z = 0.0146$ (dashed) and $Z = 0.0170$ (solid), and the BaSTI
  isochrones use $Z = 0.01$ (dashed) and $Z = 0.0198$ (solid) to
  bracket $Z=0.0161$.  \label{mandr}}
\end{figure}

\begin{figure}
\vspace{-100pt}
\hspace{-80pt}
\includegraphics[scale=0.98]{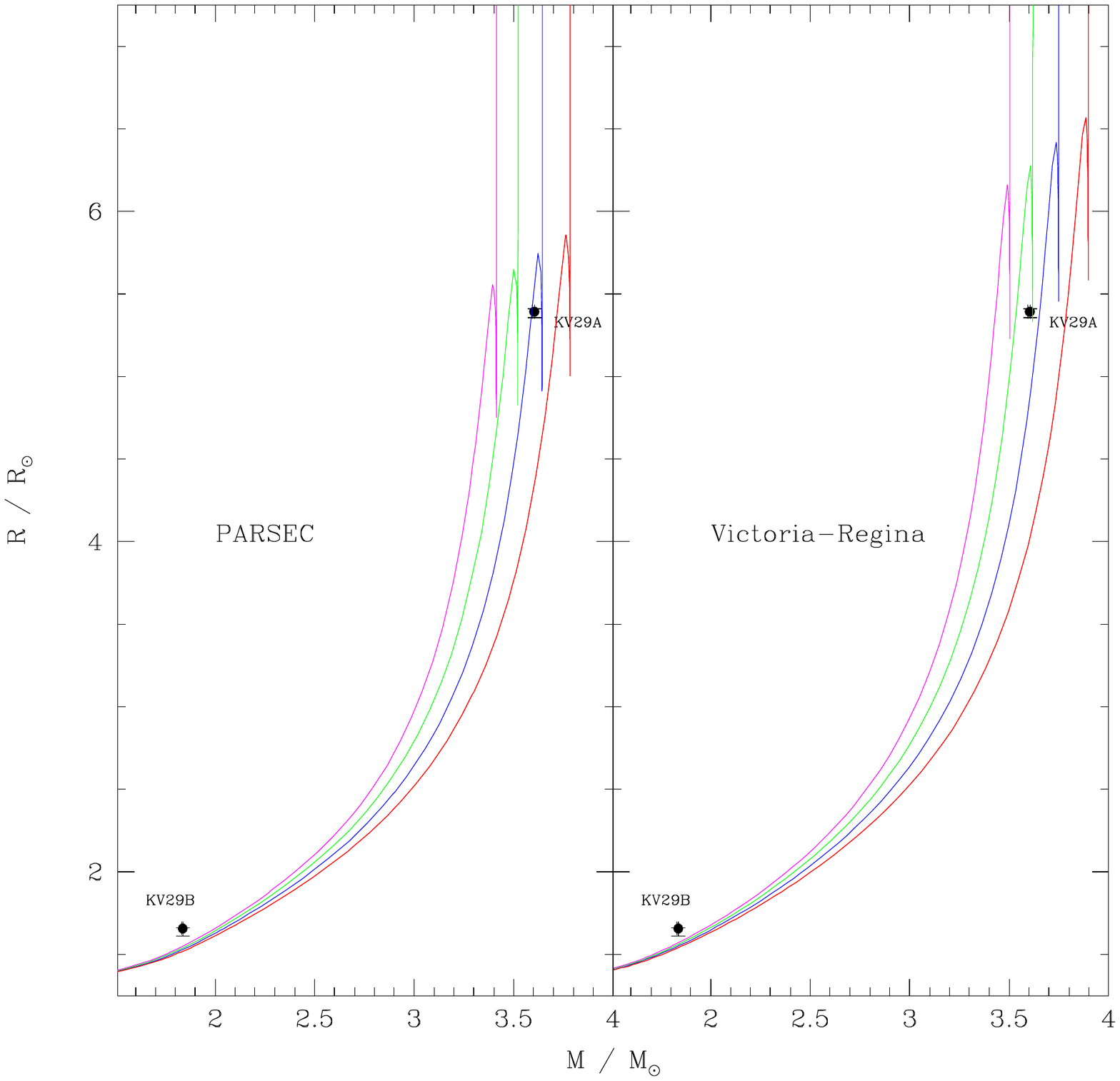}
\vspace{-150pt}
\caption{Same as mass-radius plots from Fig.\ref{mandr}, but with $Z \approx
0.0117$ to model the lower cluster metallicity of \citet{bvr14}.   The 
PARSEC and Victoria-Regina isochrones used $Z= 0.0117$ and $0.0120$ 
respectively. \label{mrbeaver}} 
\end{figure}

Fig. \ref{mandr} contains comparisons with mass versus radius ($M-R$)
isochrones. The uncertainty in KV29B's radius encompasses all four
isochrones depicted at its measured mass --- it is not sensitive to
age because it has not changed significantly in radius since the
zero-age main sequence.  While the fractional uncertainty in the
radius for KV29A is larger than that of the mass (0.6\% versus 0.3\%),
the uncertainty in the derived age is rather small because stellar
radii increase quite rapidly near the turnoff, so that mass becomes
the more important quantity for determining age.  Using the mass and
radius of KV29A, the PARSEC isochrones return an age for M11 of
$222^{+2}_{-3}\pm5$ Myr.  The first set of quoted uncertainties are
from the statistical measurement uncertainties for the mass and
radius, but systematic errors are currently a larger contributor to
the age uncertainty at present.  As discussed earlier, \citet{bvr14}
determined a lower metallicity for M11 ([Fe/H]$=-0.06\pm0.05$) using
Str\"{o}mgren photometry, and their [Fe/H] along with the
\citeauthor{asp09} solar abundance implies $Z = 0.0117\pm0.0015$.  The
lower metallicity PARSEC models yield a younger age of 221 Myr, and
this is the basis of the quoted $\pm5$ Myr uncertainty. The $M-R$
comparisons for lower metallicity isochrones in the PARSEC and
Victoria-Regina sets are shown in Fig. \ref{mrbeaver}.  For the
Victoria-Regina isochrones in Fig. \ref{mrbeaver}, we displayed tracks
for $Z = 0.0120$ only, as it is very close to the target value of $Z
= 0.0117$.  The prefered $Z$ still lies in the range chosen for the
BaSTI model of Fig. \ref{mandr}, and shows a younger age near 215 Myr
by the same method. We note that a higher $Z$ value (possibly
  because the \citealt{asp09} value for $Z_\odot$ is underestmated)
  would improve the agreement between the models and the $M-R$
  combination for the secondary star and would increase the measured age.

Differences in the physics incorporated in different sets of model
isochrones can also produce age uncertainties, and we estimate the
size of these uncertainties by examining Victoria-Regina and BaSTI
isochrones.  Interpolating between the Victoria-Regina isochrones with
$Z = 0.0146$ and $Z = 0.0170$, we find an age near 235
Myr. Interpolating between BaSTI models with $Z = 0.01$ and $Z =
0.0198$, we find an age near 215 Myr.  Based on these factors, we
estimate a systematic error due to model physics of approximately
$\pm10$ Myr.

Previous work on M11 has derived ages between 250 and 320 Myr ---
almost a 30\% range in ages. Our preferred age is approximately 9\%
lower than previously determined by \citet{sun99} but marginally
consistent. The measurement uncertainty in the age is less than 4\%,
but it must be remembered that this is model dependent and rests on
the validity of the chemical composition and physics used in the
stellar models. One of the most important physics issues is convective
core overshooting because it critically affects the amount of hydrogen
that is burned during the main sequence phase --- more overshooting
brings more fuel into the core, delays core hydrogen exhaustion, and
produces younger looking stars. BaSTI isochrones employ a larger
amount of convective core overshooting than the other models (which
affects core hydrogen exhaustion and the ``kink'' at the cluster
turnoff in the isochrones in the CMD).

\begin{figure}
\vspace{-40pt}
\hspace{-170pt}
\includegraphics[scale=1.00, angle=90]{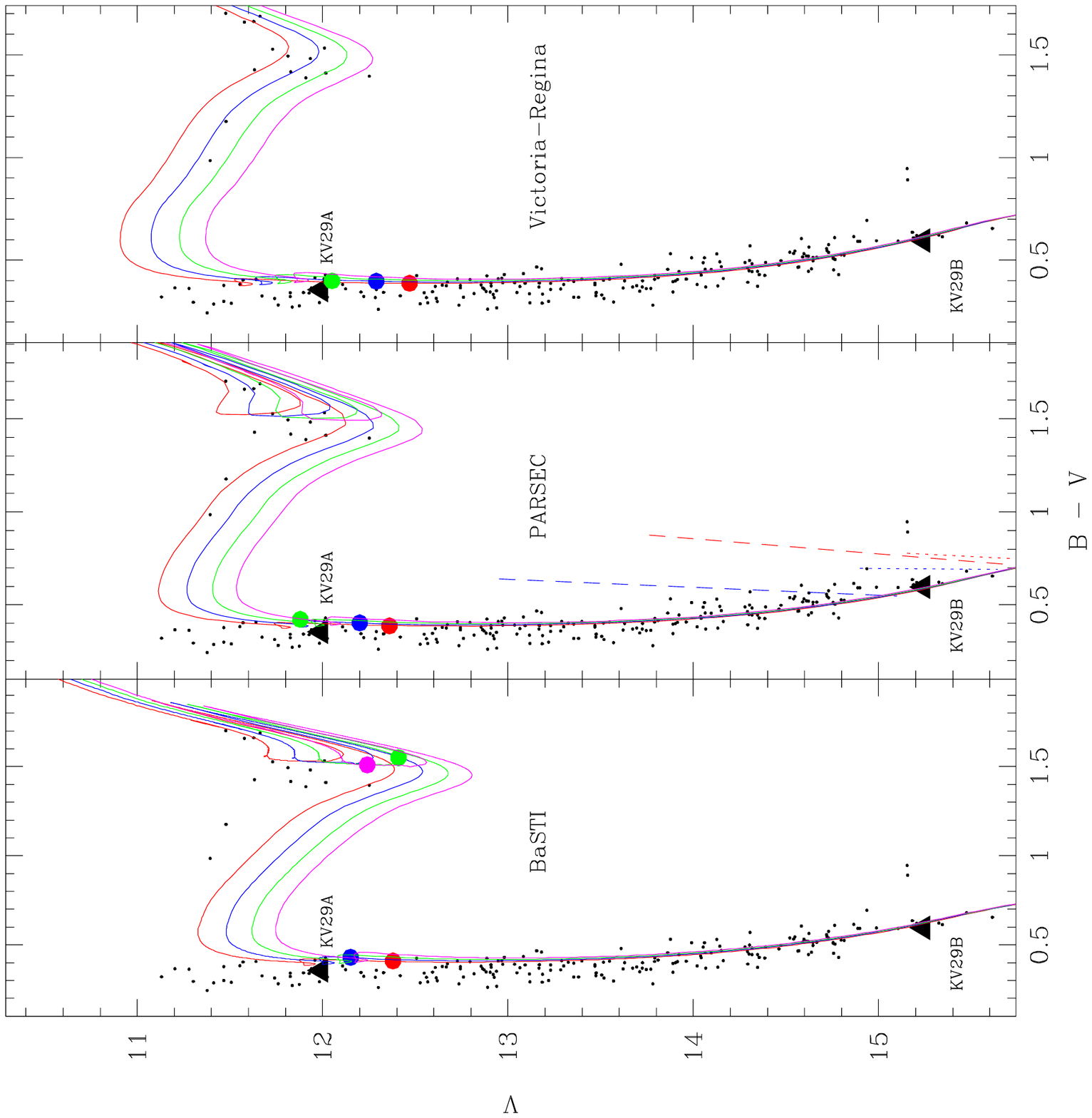}
\vspace{-30pt}
\caption{Color-magnitude diagrams for M11 (photometry from
  \citealt{ste00}).  The photometry of the system and of each
  component are shown with triangles.  Models have ages 200 (red), 220
  (blue), 240 (green), and 260 Myr (magenta), and have been shifted so
  that the isochrones match the secondary star photometry at its
  measured mass. The corresponding mass of the isochrone to the
  primary star is shown as a large circle of the same color as the
  isochrone. The BaSTI, PARSEC, and Victoria-Regina
  isochrones use $Z = 0.0198$, $0.0161$, and $0.0146$
  respectively. $\delta$ Sct (long dashed) and $\gamma$ Dor
 (short dashed) instability strips \citep{uytter} are shown in the central panel. \label{decomp}}
\end{figure}

The primary reason there has been such a large amount of
disagreement among earlier age studies is that there is little reason
to prefer an age based on isochrone shape alone.  With a combination
of the masses and photometry we bring to bear here, it is unnecessary
to fit isochrone shape.  In Section 3.2, we derived the photometry for
the two stars using the $BV$ photometry for KV29 and luminosity ratios
for each filter band from the binary star models.  In
Fig. \ref{decomp}, we plot each component on a CMD.  The photometry for
the CMDs was taken from \citet{ste00}, and the stars chosen for the
CMD are high probability members of M11 based on proper motions from
\citet{mcn77}. Each isochrone was shifted to match the secondary
star's mass and photometry simultaneously. The secondary star is an
unevolved main sequence star, and its position should be relatively
insensitive to physics in the stellar models. The primary star's mass
and photometry then constrain the cluster's age, as do the CMD
positions of upper main sequence stars. In Fig. \ref{decomp}, the
predicted position of the primary star (given its measured mass) is
marked as a circle matching the color of the isochrone for each
age. (For some ages, a star with the primary's mass would have already
evolved off the main sequence, and is not visible in the
figure.) Roughly speaking, each isochrone set predicts a CMD position
for the primary star that is in rough agreement with what is expected
from the $M-R$ diagrams discussed above.

For young clusters like M11, the lack of subgiant stars makes it
difficult to precisely identify where the turnoff lies. None of the
isochrone sets does a particularly good job of matching the
characteristics of the upper main sequence, although an unidentified
population of unresolved binaries there could be influencing that
comparison. Binary stars on the main
sequence obscure where single stars terminate in the CMD: the
combined light of a brighter MS star at the turnoff with a fainter
companion can displace an unresolved binary vertically in the CMD.
The theoretical models all predict that there is a large acceleration
in the rate of radius change at the cluster turnoff where there is a
small kink in the CMD. However, there are seemingly main sequence stars
brighter than this point in the CMD, as can be seen in
Fig. \ref{decomp}. Based on the expectations for a rapid decrease in
the evolutionary timescale after central hydrogen exhaustion and the
observed lack of stars on the subgiant branch, there should be few or
no stars on the main sequence brighter than this kink.  This leads to
the prediction that the stars brighter than KV29A will be found to
be either blue stragglers or unidentified binaries after more detailed
examination. The fact that the brightest of the main sequence stars is
a little more than 0.75 mag brighter than KV29A provides some
additional evidence that KV29A may be a star that is very close to
central hydrogen exhaustion.

On a final note, the giant clump tends to
be too bright in the PARSEC models compared to the
observations. However, this is not necessarily a way of distinguishing
between metallicities because of the effects that different physics
(like convective mixing length) and color-temperature transformations
can have.

There are several ways in which the isochrone comparisons can be
improved with future work.  Along with KV29, \citet{koo07} found two
other detached binary candidates near the turnoff (KV35 and KV36), and
there may be others. Characterization of systems like these will
improve the precision of the age comparison and can also provide
constraints on chemical composition variables like the
difficult-to-measure helium abundance \citep{bro}.  Both systems are
more challenging to analyze fully: KV35 has strong spot modulation in
its light curve, and KV36 has a period very near 12 sidereal days.  We
currently have both systems under study, but new discoveries would
help more clearly establish the cluster age by allowing us to check
the fidelity of the isochrones and the physics that goes into them.
In addition, radial velocity information on the brightest cluster
members \citep[e.g.][]{mathm11} would help clean the CMD of unresolved
binaries that conceal the position of the cluster evolutionary
sequence, especially at the bright end of the main sequence and
subgiant branch. \citet{cg14} presented radial velocities for a large
sample of cluster stars, and find several stars brighter than KV29
that are radial velocity members, although these were not monitored
for a long enough period to be certain that they are single.  Stars
that could be validated as single cluster members would produce a
significant constraint on the age. Although there are cluster members
with colors between the main sequence and red clump, the evidence so
far is that these are binary stars. Two stars with $V\approx 11.4$ and
$1.0 \la B-V \la 1.2$ (identifiers MPS 926 and 1223 in
\citealt{mcn77}) are single-lined spectroscopic binaries that each
probably contain a giant and a main sequence star, while another (MPS
1364) has a velocity that is slightly offset from the cluster mean,
possibly indicating binarity \citep{leeeb}. We identified one other
likely cluster member (MPS 670) that is closer to, but redder than,
the main sequence and should be studied to identify whether it is a
binary or a subgiant.

As a final task, we can calculate a luminosity for each star in KV29
using the photometric effective temperatures and radii from the binary
analysis.  With $R_1 = 5.39^{+0.02}_{-0.04} R_{\sun}$ and $T_{\rm
  eff,1} = 9480\pm550$ K, the primary star has a luminosity of
$212\pm49 L_{\sun}$.  The secondary star, with $R_2
=1.66^{+0.01}_{-0.04}R_{\sun}$ and $T_{\rm eff,2} = 7810\pm480$ K, has
a luminosity of $9.2\pm2.3 L_{\sun}$.  Using a bolometric correction
($BC = -0.13$; \citealt{flo96}) along with $R$ and $T_{\rm eff}$, we
can compute the distance modulus $(m-M)_V$. We find $M_{V1} = -0.93$
and $(m-M)_V = 12.90\pm0.25$ from the primary star.  Distance modulus
calculations using the secondary star ($BC = 0.03$; \citealt{flo96})
give $(m-M)_V =12.91\pm0.27$. In both cases, the uncertainty is
dominated by the temperature uncertainty, but we have included
contributions from the radius and bolometric corrections (probably at
the level of a few centimag: for example, values from \citet{vbc} are
$-0.15$ and 0.02 for the two stars). So, the binary star measurements
are very consistent with each other, giving a weighted average of
$(m-M)_V = 12.91\pm0.18$. This is consistent with the dereddened
distance modulus given by \citet{sun99} of $(m-M)_0 = 11.55\pm0.10$ if
we use their reddening and a traditional extinction factor $R_V = A_V
/ E(B-V) = 3.1$.

\section{Conclusion}

We have presented a study of the characteristics of the stars in the
eclipsing binary KV29 that is a member of the open cluster M11 and
is found at the cluster turnoff.  The DEB follows a nearly circular
orbit with a period of 4.6428 d, and it shows a total eclipse of the
secondary star.  The primary star is significantly evolved, possessing
a radius that is significantly larger than a main sequence star of the
same mass, which puts the star at the turnoff of the CMD for the
cluster. We have derived an age of $222^{+2}_{-3}\pm15$ Myr for M11
using isochrone models with the most up-to-date physics inputs with
uncertainties stemming from errors in the calculated masses and radii,
and the systematic uncertainities from the metallicity of the open
cluster and differences in the physics used in different isochrone
models.  We have also determined a distance modulus $(m-M)_V = 12.91
\pm0.18$, incorporating uncertainties in the
extinction, metallicity, and temperature.

The stars in the binary appear to be rotating close to synchronism,
despite the relatively rapid radius evolution of the primary about
over the last $\sim 40 Myr$. The binary appears to have fairly strong
tidal interactions now, and is clearly destined for strong
interactions in the near future. Models indicate that the primary star
will expand up to $6-6.5R_{\sun}$ over the next 15 Myr. After a short
contraction back to near its present size, it will rapidly expand and
overflow its Roche lobe shortly after beginning to transit the
Hertzsprung gap on a thermal timescale. Strong mass transfer will be
aided by shrinkage of the orbital separation and of the primary's
Roche lobe, and will make it impossible for the lower-mass secondary
star to accomodate the donated gas, leading to a contact phase. After
the system's mass ratio has been reversed, the current primary star
will eventually complete its mass transfer, and the system will be
fairly quiescent until the current secondary star evolves and expands
and starts an Algol mass-transfer phase in the system.

The $BVR$ light curves for KV29 showed significant amounts of variation,
making it challenging to precisely determine the
radii of the two components.  We binned the lightcurves in phase in
order to smooth these variations.  We find uncertainties in the
primary and secondary radii of 0.6\% and 2.7\%, respectively. In spite of
these errors, the age uncertainty is rather low due to the evolved
nature of the primary star.  The age of the primary is very sensitive
to the value of its mass \citep{sou04}, and as a result we have
improved the precision of the cluster age over CMD-based methods that
are subject to more substantial systematic errors related to interstellar
extinction. Our age determination is lower than all previous CMD-based
determinations, and has considerably higher precision.

With the precision of M11's age, it can be put confidently
in perspective relative to other clusters, and this allows calibration
of other age indicators. Stellar rotation \citep{barnes07} is an
important age indicator because it can potentially be applied to
individual field stars and it retains sensitivity to age on the main
sequence when many other characteristics of stars are changing
minimally. The primary limitation of gyrochronology is the need for
calibrators of known age, and open clusters like M11 provide
that. When the calibrating clusters have precisely known ages, there
is no ambiguity as to the relative ages of different clusters. With an
age uncertainty of a few percent, the rotation periods measured for
M11 stars \citep{messina} as a function of mass (or rather, its proxy,
color) can be reliably compared with clusters of similar age like M35
\citep{mei09}, M34 \citep{mei11}, and M37 \citep{mess08}.

The photometry of the secondary star in the eclipsing binary appears
to put it in the instability strip where it crosses the main
sequence. This part of the instability strip is inhabited by $\delta$
Scuti and $\gamma$ Doradus pulsating stars, and more than a 
dozen $\delta$ Sct stars have previously been detected via their
photometric variability \citep{har05,koo07,lee10}. Seven $\delta$ Sct
are likely cluster members according to proper motions \citep{mcn77},
but these are likely to be just the stars with the highest pulsation
amplitudes. With higher precision photometry, the pulsation
characteristics of the stars in this cluster ensemble may enable an
independent determination of the cluster distance modulus. The
measured mass of the secondary star could also help directly constrain
the properties of the pulsators in M11 and produce stronger tests of
how well we understand their pulsation spectra.

The measurement precision on the characteristics of KV29 will improve
with additional spectroscopic and photometric eclipse observations
that allow us to reduce the influence of the observed variability
  in the light curve shape. Improved determination of the effective
temperatures would improve the distance modulus and the spectral
typing of both components, and would have some effect on the eclipse
modelling through the limb darkening prescription. Photometric data in
additional colors (like $I$ band) and improved spectroscopic
constraints would improve this. As mentioned before, we are in the
process of investigating other DEBs in M11 that are known to display
eclipses. With the high-precision study of multiple eclipsing systems
in the cluster, we will be able to statistically hone the cluster age.

\acknowledgements

This work has been funded through grant AST 09-08536 from the
National Science Foundation to E.L.S. We would like to thank the Director of
Mount Laguna Observatory (P. Etzel) for generous allocations of observing
time, and K. Brogaard for the use of his spectral disentangling code.

The Hobby-Eberly Telescope (HET) is a joint project of the University of Texas
at Austin, the Pennsylvania State University, Stanford University,
Ludwig-Maximilians-Universitat Munchen, and Georg-August-Universitat
Gottingen.  The HET is named in honor of its principal benefactors, William
P. Hobby and Robert E. Eberly.  

This research made use of the SIMBAD database, operated at CDS, Strasbourg, 
France; the NASA/IPAC Infrared Science Archive, which is operated by the 
Jet Propulsion Laboratory, California Institute of Technology, under contract 
with the National Aeronautics and Space Administration; and the WEBDA database, 
operated at the Institute for Astronomy of the University of Vienna.

\begin{figure}
\vspace{-40pt}
\hspace{-170pt}
\includegraphics[scale=1.00, angle=90]{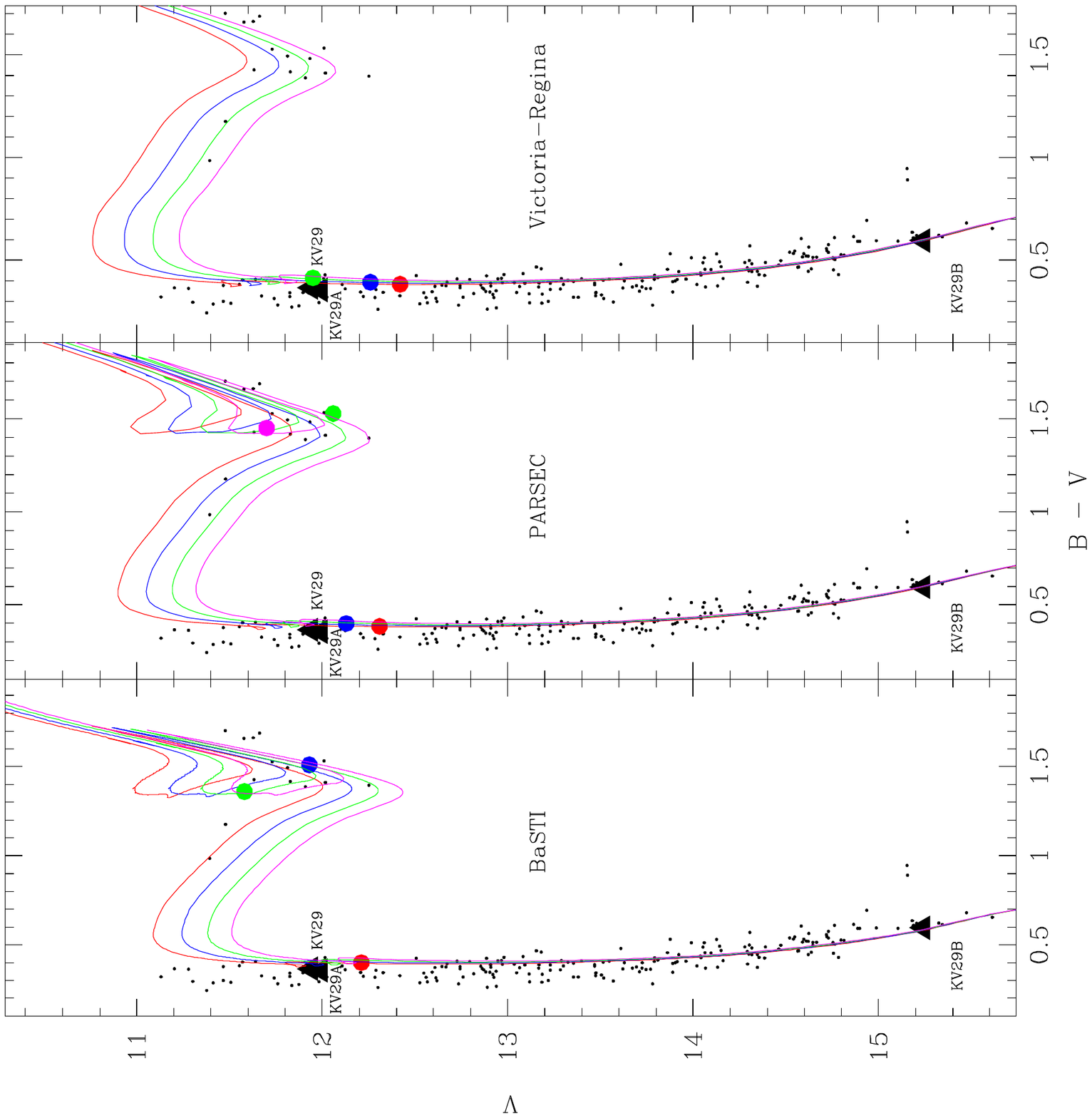}
\caption{Same as Fig.\ref{decomp} except with the $Z \approx 0.0117$
  \citep{bvr14}.  The BaSTI, PARSEC, and Victoria-Regina isochrones
  use $Z= 0.01$, $ 0.0117$, and $0.0120$,
  respectively.\label{decompbeaver}}
\end{figure}

\begin{deluxetable}{ccccc}
\tabletypesize{\scriptsize}
\tablecaption{Log of Photometric Observations and Data Points}
\tablewidth{0pt}
\tablehead{
\colhead{Civil Date} & \colhead{HJD$-2450000$} & \colhead{$N_B$} & \colhead{$N_V$} & \colhead{$N_R$}
}
\startdata
28 June 2009 & 5011 & & 131 & \\
29 June 2009 & 5012 & & 83 & \\
30 June 2009 & 5013 & & 48 & \\
1 June 2010 & 5349 & & 111 & \\
2 June 2010 & 5350 & 114 & & \\
10 June 2010 & 5358 & & 111 & \\
17 June 2010 & 5365 & & 124 & \\
25 June 2010 & 5373 & & 21 & \\
2 July 2010 & 5380 & & 146 & \\
9 July 2010 & 5387 & & 104 & \\
16 July 2010 & 5394 & 130 & & \\
23 July 2010 & 5401 & 133 & & \\
30 July 2010 & 5408 & 22 & & \\
1 August 2010 & 5410 & 42 & 45 & \\
8 August 2010 & 5417 & 52 & 52 & \\
15 August 2010 & 5424 & 41 & 42 & \\
18 August 2010 & 5427 & & & 27 \\
1 September 2010 & 5441 & 24 & 24 & \\
6 September 2010 & 5446 & 23 & 23 & \\
7 September 2010 & 5447 & 21 & & 20 \\
10 September 2010 & 5450 & 14 & 15 & \\
14 September 2010 & 5454 & 21 & 21 & 19 \\
17 September 2010 & 5457 & 17 & 17 & 17 \\
19 September 2010 & 5459 &  & & 82 \\
21 September 2010 & 5461 & 11 & 14 & 15 \\
26 September 2010 & 5466 & & & 74 \\
27 May 2011 & 5709 & & 60 & \\
22 June 2011 & 5735 & & & 150 \\
29 June 2011 & 5742 & & & 79 \\
6 July 2011 & 5749 & & & 151 \\
13 July 2011 & 5756 & & & 147 \\
20 July 2011 & 5763 & 83 & & \\
7 August 2011 & 5781 & 40 & &  \\
13 August 2011 & 5787 & 37 & 36 & 34 \\
18 August 2011 & 5792 & 25 & & 26 \\
27 May 2013 & 6439 & & & 59\\
\tableline
\enddata
\label{photobs}
\end{deluxetable}

\begin{deluxetable}{cccrr}
\tabletypesize{\scriptsize}
\tablecaption{HET Spectroscopic Observations}
\tablewidth{0pt}
\tablehead{\colhead{UT Date} & \colhead{HJD$-2450000$} & \colhead{Phase} & \colhead{$V_1$ (km s$^{-1}$)\tablenotemark{a}} & \colhead{$V_2$ (km s$^{-1}$)\tablenotemark{a}}}
\startdata
6 May 2010 & 5322.9402 & 0.049 & $12.39 \pm 1.54$ & $78.29 \pm 2.08$ \\
6 May 2010 & 5322.9461 & 0.050 & $11.09 \pm 1.23$ & $81.19 \pm 1.49$ \\
15 June 2010 & 5362.8405 & 0.643 & $92.27 \pm 1.08$ & $-79.13 \pm 5.96$ \\
25 June 2010 & 5372.8164 & 0.792 & $105.03 \pm 1.85$ & $-104.92 \pm 1.49$ \\
28 July 2010 & 5405.7272 & 0.880 & $89.00 \pm 2.47$ & $-64.00 \pm 1.19$ \\
1 August 2010 & 5409.7096 & 0.738 & $110.68 \pm 6.32$ & $-111.87 \pm 2.38$ \\
13 August 2010 & 5421.6892 & 0.318 & $-34.97 \pm 2.00 $ & $168.13 \pm 0.60$ \\
5 September 2010 & 5444.6364 & 0.261 & $-40.83 \pm 3.85$ & $182.52 \pm 1.19$ \\
12 September 2010 & 5451.6024 & 0.761 & $105.58\pm 1.54$ & $-108.32 \pm 1.19$ \\
19 September 2010 & 5458.5978 & 0.268 & $-41.29 \pm 2.47$ & - \\
24 August 2011 & 5797.6628 & 0.297 & $-36.37 \pm 0.77$ & $179.38 \pm 2.98$ \\
29 August 2011 & 5802.6365 & 0.368 & $-22.13 \pm 1.54$ & $143.62 \pm 0.89$ \\
7 September 2011 & 5811.6120 & 0.301 & $-36.32 \pm 1.54$ & $169.78 \pm 5.66$ \\
20 September 2011 & 5824.5937 & 0.097 & $-5.97 \pm 2.00$ & $116.13 \pm 3.57$ \\
21 September 2011 & 5825.6048 & 0.315 & $-33.89 \pm 2.31$ & $168.41 \pm 2.98$ \\
22 September 2011 & 5826.6063 & 0.531 & $45.96 \pm 4.78$ & - \\
29 September 2011 & 5833.5679 & 0.030 & $20.07 \pm 4.16$ & $59.82 \pm 0.89$ \\
\tableline
\enddata
\tablenotetext{a}{Errors are scaled to return a reduced $\chi^2 \approx 1$}
\label{specobs}
\end{deluxetable}

\begin{deluxetable}{cc}
\tabletypesize{\scriptsize}
\tablecaption{Constrained Values for KV29}
\tablewidth{0pt}
\tablehead{\colhead{Parameter} & \colhead{Value}}
\startdata
$Q=M_2/M_1$ & $0.509 \pm 0.004$  \\
$K_{p}$ (km s$^{-1}$) & $74.25 \pm 0.50$   \\
$e$ & $0.0$ \\
$T_1$ (K) & $9480 \pm 550$ \\
$\gamma$ (km s$^{-1}$) & $32.39 \pm 1.90$  \\
$x_{R,1}$ & 0.2710  \\
$y_{R,1}$ & 0.4228  \\
$x_{R,2}$ & 0.3504  \\
$y_{R,2}$ & 0.3981  \\
$x_{V,1}$ & 0.2159   \\
$y_{V,1}$ & 0.3824  \\
$x_{V,2}$ & 0.2765  \\
$y_{V,2}$ & 0.3764   \\
$x_{B,1}$ & 0.1634  \\
$y_{B,1}$ & 0.3416   \\
$x_{B,2}$ & 0.1947  \\
$y_{B,2}$ & 0.1241  \\
\enddata
\label{constraints}
\end{deluxetable}

\begin{deluxetable}{ccc}
\tabletypesize{\scriptsize}
\tablecaption{Best-Fit Model Parameters for KV29}
\tablewidth{0pt}
\tablehead{\colhead{Parameter} & \colhead{Constant LD \tablenotemark{a}} & 
\colhead{Fitted LD\tablenotemark{b}}}
\startdata
$i$ ($\degr$) & $78.48^{+0.30}_{-0.06}$ & $78.38^{+0.37}_{-0.25}$ \\
$P$ (d) & $4.64276\pm0.00001$ & $4.64287\pm0.00006$ \\
$t_o$ (HJD) & $2455373.788\pm0.001$ & $2455373.790\pm0.001$ \\
$R_1/a$ & $0.2620^{+0.0008}_{-0.0014}$ & $0.267\pm0.002$ \\
$R_2/a$ & $0.0804^{+0.0003}_{-0.0021}$ & $0.078\pm0.002$ \\
$R_1/R_2$ & $3.256^{+0.074}_{-0.011}$ & $3.428^{+0.089}_{-0.067}$ \\
$(R_1+R_2)/a$ & $0.342^{+0.001}_{-0.003}$ & $0.345^{+0.003}_{-0.004}$ \\
$T_2/T_1$ & $0.840\pm0.002$ & $0.866\pm0.005$ \\
$a / R_\sun$ & $20.591^{+0.004}_{-0.022}$ & $20.608^{+0.017}_{-0.007}$ \\
$a \sin i / R_\sun$ & $20.176\pm0.021$ & $20.17652\pm0.00006$\\
$M_1/M_{\sun}$ & $3.604^{+0.002}_{-0.011}$  & $3.613^{+0.009}_{-0.004}$ \\
$M_2/M_{\sun}$ & $1.837^{+0.001}_{-0.006}$ & $1.842^{+0.005}_{-0.002}$ \\
$R_1/R_{\sun}$ & $5.392^{+0.018}_{-0.035}$ & $5.407^{+0.001}_{-0.040}$ \\
$R_2/R_{\sun}$& $1.656^{+0.007}_{-0.044}$ & $1.699^{+0.052}_{-0.001}$ \\
log $g_1$ (cgs) & $3.531^{+0.004}_{-0.003}$ & $3.530^{+0.006}_{-0.0}$ \\
log $g_2$ (cgs) & $4.264^{+0.022}_{-0.004}$ & $4.243^{+0.0}_{-0.025}$ \\
$M_1 \sin^3 i$ ($M_\odot$) & $3.391\pm0.011$ & $3.391\pm0.003$ \\
$M_2 \sin^3 i$ ($M_\odot$) & $1.728\pm0.005$ & $1.7283\pm0.0005$ \\
$V_{1,rot} \sin i$ (km s$^{-1}$)\tablenotemark{c} & $57.60^{+0.18}_{-0.32}$ & $58.30^{+0.35}_{-0.46}$ \\
$V_{2,rot} \sin i$ (km s$^{-1}$)\tablenotemark{c} & $17.69^{+0.07}_{-0.46}$ & $17.01^{+0.39}_{-0.52}$ \\
\tableline
\enddata
\tablenotetext{a}{Used all photometry. Prefered fit.}
\tablenotetext{b}{Used phase-binned light curves.}
\tablenotetext{c}{Calculated assuming synchronous rotation.}
\label{trials}
\end{deluxetable}

\begin{deluxetable}{cccc}
\hspace{-50pt}
\tabletypesize{\scriptsize}
\tablecaption{Deconvolved Photometry of KV29}
\tablewidth{0pt}
\tablehead{\colhead{Name} & \colhead{$V$} & \colhead{$B-V$}}
\startdata
KV29 & $11.921\pm0.004$\tablenotemark{a} & $0.392\pm0.008$\tablenotemark{a} \\ 
KV29A & $11.974\pm0.005$ & $0.382\pm0.008$ \\ 
KV29B & $15.226\pm0.015$ & $0.624\pm0.023$ \\
\tableline
\enddata
\tablenotetext{a}{\citet{ste00}}
\label{colors}
\end{deluxetable}
\end{document}